\documentclass[aps, prl, tightenlines, twoside, secnumarabic, twocolumn,
superscriptaddress, showpacs, preprintnumbers, nofootinbib, notitlepage, fleqn]{revtex4-1}

\pdfoutput=1
\usepackage[english]{babel}
\usepackage{amsmath,amssymb,amsfonts, bm,bbm,slashed}
\usepackage{graphicx}
\usepackage[sort&compress]{natbib}
\usepackage{xcolor}
\usepackage[normalem]{ulem}
\usepackage{hyperref}
\usepackage{cleveref}
\definecolor{red}{rgb}{1.0, 0, 0}
\usepackage{hyperref}
\usepackage{enumerate}
\usepackage{epsfig, subfigure}
\usepackage{setspace}
\usepackage{booktabs, tabularx}
\usepackage{units}
\usepackage[utf8]{inputenc}
\usepackage{lineno}

\hypersetup{
    pdfnewwindow=true,   
    colorlinks=true,     
    linkcolor=blue,      
    citecolor=blue,      
    filecolor=blue,      
    urlcolor=blue        
}

\allowdisplaybreaks

\bibpunct{[}{]}{,}{n}{}{,}

\newcommand{\ev}[1]{\ensuremath{\left\langle #1 %
                     \right\rangle}} 

\newcommand{\vev}[1]{{\langle #1 \rangle}}

\begin{document}

\title{The Vev Flip-Flop: Dark Matter Decay between Weak Scale Phase Transitions}

\author{Michael J.\ Baker}
\email{micbaker@uni-mainz.de}
\affiliation{PRISMA Cluster of Excellence \& Mainz Institute for Theoretical Physics,
             Johannes Gutenberg University, Staudingerweg 7, 55099 Mainz, Germany}

\author{Joachim Kopp}
\email{jkopp@uni-mainz.de}
\affiliation{PRISMA Cluster of Excellence \& Mainz Institute for Theoretical Physics,
             Johannes Gutenberg University, Staudingerweg 7, 55099 Mainz, Germany}

\date{\today}
\preprint{MITP/16-090}

\begin{abstract}
  We propose a new alternative to the Weakly Interacting Massive Particle
  (WIMP) paradigm for dark matter.  Rather than being determined by thermal
  freeze-out, the dark matter abundance in this scenario is set by dark matter
  decay, which is allowed for a limited amount of time just before the
  electroweak phase transition.  More specifically, we consider fermionic singlet
  dark matter particles coupled weakly to a scalar mediator $S_3$ and to auxiliary dark sector
  fields, charged under the Standard Model gauge groups.  Dark
  matter freezes out while still relativistic, so its abundance is initially
  very large. As the Universe cools down, the scalar mediator develops a vacuum
  expectation value (vev), which breaks the symmetry that
  stabilises dark matter. This allows dark matter to mix with charged fermions and
  decay.  During this epoch, the dark matter abundance is reduced to give the
  value observed today.  Later, the SM Higgs field also develops a vev, which
  feeds back into the $S_3$ potential and restores the dark sector symmetry.
  In a concrete model we show that this  ``vev flip-flop'' scenario is
  phenomenologically successful in the most interesting regions of its
  parameter space. We also comment on detection prospects at the LHC and
  elsewhere.
\end{abstract}

\maketitle

The WIMP (Weakly Interacting Massive Particle) paradigm in dark matter physics
states that dark matter (DM) particles should have non-negligible couplings to
Standard Model (SM) particles.  Their abundance today would then be determined by their
abundance at freeze-out, the time when the temperature of the Universe dropped
to the level where interactions producing and annihilating DM particles become
inefficient.  In many models these interactions should still be observable today,
through residual DM annihilation in galaxies and galaxy clusters, through
scattering of DM particles on atomic nuclei, or through DM production at colliders.
The conspicuous absence of any convincing signals~\cite{LUX:2016, Tan:2016zwf,
Ackermann:2015zua, Pizzolotto:2016lwk, Madhavacheril:2013cna, Aaboud:2016tnv,
CMS:2016pod} to date motivates us to look for
alternatives to the WIMP paradigm.

In this letter, we present a scenario in which the DM abundance is set not by
annihilation, but by decay.  We focus on models in which fermionic DM particles
couple to a new scalar species and argue that the resulting scalar potential
may undergo multiple phase transitions, ``flip-flopping'' between phases in
which the symmetry stabilising the DM is intact and a phase
where it is broken. Similar behaviour is found in models of electroweak
baryogenesis~\cite{Profumo:2007wc, Cline:2009sn, Espinosa:2011ax, 
Cui:2011qe, Cline:2012hg, Fairbairn:2013uta, Curtin:2014jma},
see also \cite{Cohen:2008nb} for related work.
During the broken phase, the initially overabundant DM
particles are depleted until their relic density reaches the value observed
today.  Although the resulting picture of early freeze out, followed  by a
period of DM decay around the weak scale, is quite generic, we focus here on a
concrete example and comment on possible generalisations in the end.
\newline


\begin{table}
  \centering
  \begin{ruledtabular}
  \begin{tabular}{ccccc}
    Field     &      Spin      &     SM      & $\mathbb{Z}_3$ & mass scale \\\hline
    $\chi$    & $\tfrac{1}{2}$ & $(1, 1, 0)$ & $\chi \to e^{2\pi i/3} \chi$        &   TeV   \\
    $S_3$     &       $0$      & $(1, 3, 0)$ & $S_3 \to e^{2\pi i/3} S_3$          & 100~GeV \\
    $\Psi_3$  & $\tfrac{1}{2}$ & $(1, 3, 0)$ & $\Psi_3 \to e^{-2\pi i/3} \Psi_3$   &   TeV   \\
    $\Psi_3'$ & $\tfrac{1}{2}$ & $(1, 3, 0)$ & $\Psi_3' \to e^{-2\pi i/3} \Psi_3'$ &   TeV   \\
  \end{tabular}
  \end{ruledtabular}
  \caption{New particles in our toy model, with their mass scales and charge
    assignments under the SM gauge group $SU(3)_c \times SU(2)_L \times U(1)_Y$
    and under the dark sector $\mathbb{Z}_3$ symmetry.}
  \label{tab:particles}
\end{table}

\emph{Model Framework.}---We introduce a complex dark sector scalar $S_3$ that
is a triplet under the SM $SU(2)_L$ gauge symmetry and carries zero
hypercharge.  We take the DM particle to be a multi-TeV Dirac fermion $\chi$, uncharged
under the SM gauge group.  To offer the DM particle a decay mode, we also
introduce two multi-TeV Dirac fermion triplets $\Psi_3 = (\psi^+, \psi^0, \psi^-)$
and $\Psi_3' = (\psi'^+, \psi'^0, \psi'^-)$. All new
particles are charged under a symmetry that stabilises the dark sector and is
taken here as a $\mathbb{Z}_3$, which could be a remnant of a dark sector gauge
symmetry broken at a scale $\gg \text{TeV}$.  The dark sector particle content
is summarised in \cref{tab:particles}.

The tree level scalar potential and the relevant Yukawa terms in
the Lagrangian of the model are
\begin{align}
  \hspace*{-0.9cm}
  \mathcal{L}_\text{scalar}
    &= \mu_H^2 H^\dag H
     - \lambda_H (H^\dag H)^2
                              \nonumber\\[0.2cm]
    &+ \mu_S^2 S_3^\dag S_3
     - \lambda_S (S_3^\dag S_3)^2
     - \lambda_3 (S_3^\dag T^a S_3)^\dag (S_3^\dag T^a S_3)
                              \nonumber\\[0.2cm]
    &- \alpha (H^\dag H) (S_3^\dag S_3)
     - \beta (H^\dag \tau^a H) (S_3^\dag T^a S_3) \,,
                                   \label{eq:Pot} \\[0.2cm]
  \hspace*{-0.9cm}
  \mathcal{L}_\text{Yuk}
    &= y_\chi S_3^\dag \bar\chi \Psi_3
     + y_\chi' S_3^\dag \bar\chi \Psi_3'
     + y_\Psi \epsilon^{ijk} S_3^i \overline{\Psi_3^j} (\Psi_3'^k)^c
     + h.c.
                                                 \label{eq:Lag}
\end{align}
The first line in \cref{eq:Pot} contains the SM Higgs potential, characterised
by the tree-level mass parameter $\mu_H \simeq 88$~GeV and the quartic coupling
$\lambda_H \simeq 0.12$.  We will use $H = (G^+, (h+iG^0)/\sqrt{2})$.  In the
second line, the analogous potential for the scalar mediator $S_3
= (s^+, (s + i a)/\sqrt{2}, s^-)$ is given. It is
characterised by $\mu_S \sim \mathcal{O}(\text{100~GeV})$ and positive quartic
couplings $\lambda_S, \lambda_3 \sim \mathcal{O}(1)$.
We will use the convention that only the electrically neutral, CP even components
$h$ and $s$ of $H$ and $S_3$ acquire vevs.  The third line of
\cref{eq:Pot} contains the Higgs portal terms with
couplings $\alpha$ and $\beta$.  The symmetries of the
model are designed to allow new Yukawa couplings involving $S_3$,
$\chi$, $\Psi_3$ and $\Psi_3'$ with small coupling constants $y_\chi$, $y_\chi'$.
When $\vev{s} \neq 0$, these couplings will lead to mixing between $\chi$, $\Psi_3$
and $\Psi_3'$ and thus to DM decay via $\chi \to W \Psi_3$, $\chi \to W \Psi_3'$.
The masses $m_\chi$, $m_\Psi$, $m_{\Psi'}$ of the new fermions are assumed to be such that
the decay channels $\chi \to S_3 \Psi_3$, $\chi \to S_3 \Psi_3'$ are forbidden
today.  If these decay channels were open, the DM candidates would be $\psi^0$, $\psi'^0$
and $S$ instead of $\chi$, and their abundance would be determined by a regular
thermal freeze-out. DM decays to $S\Psi_3$ and $S\Psi_3'$ may still be open at early times as the masses
of $S_3$, $\Psi_3$, and $\Psi_3'$ are $T$-dependent.  In particular, when $\vev{s} \neq 0$,
the coupling $y_\Psi \sim \mathcal{O}(1)$ in the last line of \cref{eq:Lag}
leads to mass shifts for $\psi^\pm$ and $\psi'^\pm$.
The same term also allows
$\Psi_3$ and $\Psi_3'$ to annihilate efficiently into SM
particles.\footnote{It is this requirement that prompted us to introduce
  two fields $\Psi_3$ and $\Psi_3'$.  If one of them was omitted, efficient annihilation
  could not occur because a term of the form $\epsilon^{ijk} S_3^i \overline{\Psi_3}
\Psi_3^c$ vanishes.}

We assume that the parameters of the scalar potential are chosen such that, at zero
temperature, the SM Higgs field has its usual vev $\vev{h} \equiv v = 246$\,GeV,
while $\vev{s} = 0$. Thus, the electroweak symmetry is broken at temperature
$T=0$, while the $\mathbb{Z}_3$ symmetry that stabilises DM is unbroken.  The
tree level masses of the physical Higgs boson $h$ and of $s$ today are
given by $m_h^2 = 2 \mu_H^2$ and $m_s^2 = -\mu_S^2 + \alpha \mu_H^2 / (2
\lambda_H)$.
\newline


\emph{The Vev Flip-Flop.}---To determine the evolution of the system given by
\cref{eq:Lag} in the hot early Universe, we consider the effective potential
$V^\text{eff}$. In addition to $V^\text{tree}$, the effective potential
includes the $T$-independent one-loop Coleman--Weinberg
contributions~\cite{Coleman:1973jx}, the one-loop $T$-dependent
corrections~\cite{Dolan:1973qd}, and the resummed higher-order ``daisy''
contributions~\cite{Carrington:1991hz, Quiros:1999jp, Ahriche:2007jp,
Delaunay:2007wb}, see appendix~A for details. We include the dominant
contributions to these higher order terms from $t, B^\mu, Z^\mu, W^\mu, H$ and $S_3$.
The Coleman--Weinberg contribution introduces a renormalisation scale, which we
set equal to the SM Higgs vev.  Note that at high temperature $\mu_S^2$
receives negative corrections from
the one-loop $T$ dependent term, and positive corrections from the ``daisy''
terms.  This means that at high $T$, the $\mathbb{Z}_3$ symmetry may be
broken or unbroken, depending on which contribution dominates.  The SM Higgs
vev is always zero at high $T$ for the experimentally determined values
of $\mu_H^2$ and $\lambda_H$.


\begin{figure}
  \begin{center}
    \includegraphics[width=\columnwidth]{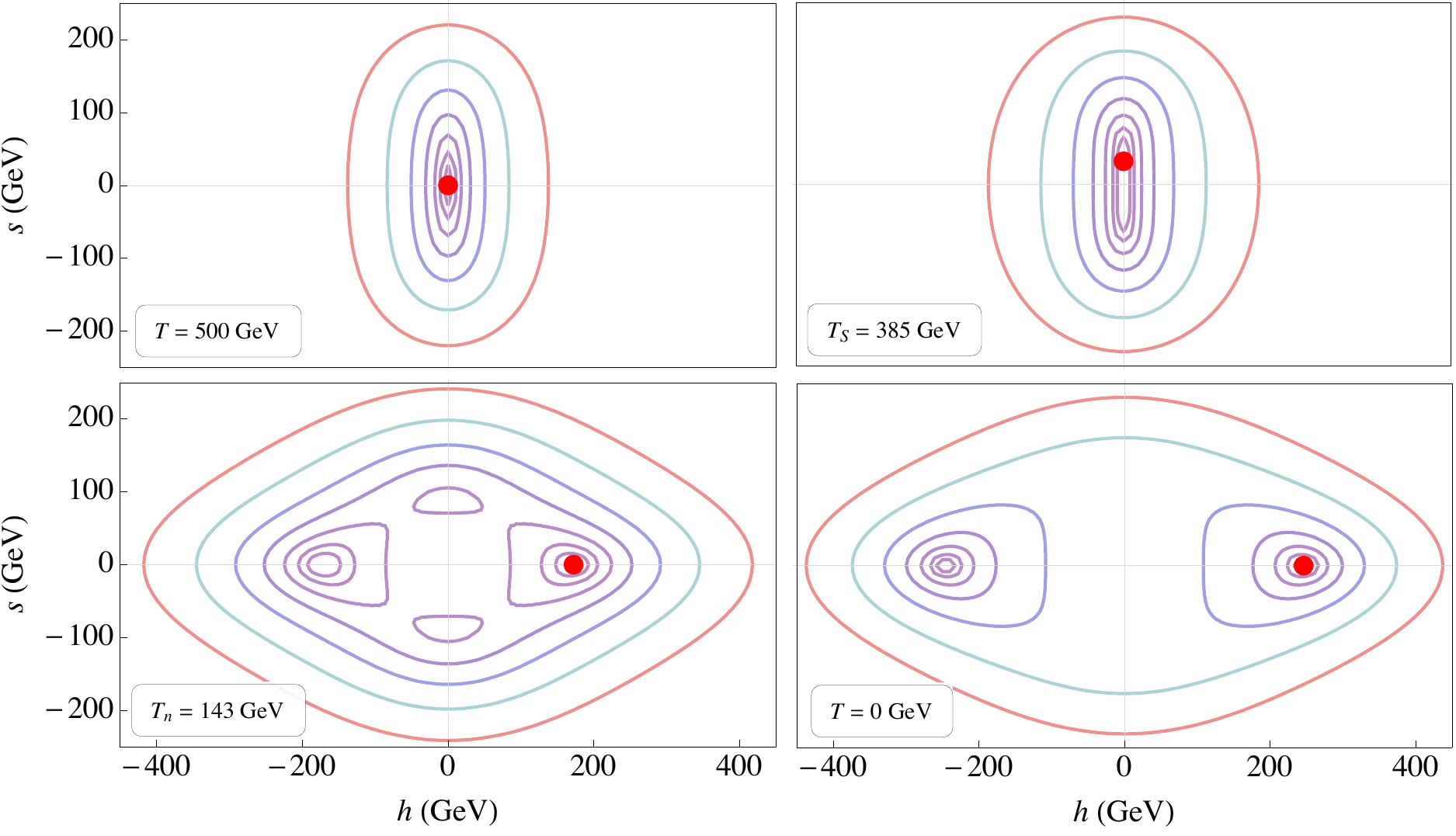}
  \end{center}
  \caption{
    The effective potential $V^\text{eff}$ at high $T$ (top left), at
    the nucleation temperature $T_S = 385$\,GeV where $S_3$ obtains a non-zero vev (top
    right), at the nucleation temperature $T_n = 143$\,GeV where the SM
    Higgs obtains a vev and the $S_3$ vev goes to zero (bottom left), and at
    $T = 0$ (bottom right).   Red dots
    indicate the minima the Universe is in at each stage.  Contour lines
    are evenly spaced on a log scale and range from $1.4 \times 10^6\text{\,GeV}^4$
    (purple) to $5.6 \times 10^8 \text{\,GeV}^4$ (red), with each contour representing
    one $e$-fold.  For this plot, we have chosen
    $\alpha= 1.33$, $\beta = 0$, $\lambda_S = \lambda_3 = 1$,
    and $\mu_S^2 > 0$ is fixed such that
    the tree level mass $m_s = 185$\,GeV at $T=0$\,GeV.}
  \label{fig:eff-pot}
\end{figure}


We illustrate the main features of the flip-flopping vevs mechanism in
\cref{fig:eff-pot} for a specific choice of Lagrangian parameters.
At high $T$ (top left panel),
$V^\text{eff}$ is symmetric in both $s$ and $h$.  As the Universe cools
the $T$-dependent corrections to the $S_3$ mass term become subdominant and,
since $\mu_S^2$ is positive, $S_3$ develops a non-zero vev at $T = T_S$, \cref{fig:eff-pot}
(top right).  Now the $\mathbb{Z}_3$ symmetry that stabilises the dark sector is broken,
and DM can decay via $\chi \to W \Psi_3$ and $\chi \to W \Psi_3'$.
Note that, during this epoch, $SU(2)_L$
is also broken and $W^\mu$ obtains a mass $m_W = g \vev{s}$, where $g$
is the SM weak gauge coupling. Thanks to the mass shift in $\psi^\pm$ and
$\psi'^\pm$, the decay modes $\chi \to S \Psi_3$ and $\chi \to S \Psi_3'$
may also be open.

As the temperature continues to drop $T$-dependent corrections to $\mu_H$
become subdominant and $V^\text{eff}$ develops new minima at non-zero $\ev{h}$.
The extra contribution of this vev to the $S_3^\dag S_3$ term through the Higgs portal
makes the effective $\mu_S^2$ negative and so $\vev{s}=0$ is restored at the
new minima. At first, the minima at $\ev{h} \neq 0$ are only local minima, but
as the temperature drops they become global minima.
Typically there is a barrier between the $\vev{s} \neq 0$ and the
$\vev{h} \neq 0$ minima, so that the phase transition between the two is first
order.  For the transition to take place, the formation and growth of bubbles
of the new phase has to be energetically favourable~\cite{Linde:1981zj}, which
happens at $T = T_n$, some time after the $\vev{h} \neq 0$ minima become the
global ones. In other words, the Universe is in a supercooled state for some
time.  To calculate the nucleation temperatures $T_S$ and $T_n$ we used the
publicly available {\tt CosmoTransitions} package~\cite{Wainwright:2011kj,
Kozaczuk:2014kva, Blinov:2015sna, Kozaczuk:2015owa}.  Note that since the
effective potential is gauge dependent away from the minima, the calculated
$T_n$ may have a residual gauge dependence, see e.g.~\cite{Patel:2011th}.  As
is usual in the baryogenesis literature, we neglect this effect.  For our
illustrative parameter point, the effective potential at $T_n$ is shown in the
bottom left panel of \cref{fig:eff-pot}.  At this point the $\mathbb{Z}_3$
symmetry is restored and $\chi$ stabilises.

The transition from a phase with unbroken $\mathbb{Z}_3$ and $SU(2)_L$ at very
high $T$ to the broken phase at intermediate $T$, and back to an unbroken phase
at the electroweak phase transition is what we call the \emph{vev flip-flop}.
As we approach the present day, $T \sim 0$, the $\ev{h} \neq 0$ minima
deepen, and the Universe remains in the phase with broken electroweak symmetry
and stable DM, \cref{fig:eff-pot} (bottom right).
\newline


\emph{Dark Matter Abundance.}---One of the salient features of the Lagrangian
in \cref{eq:Lag} is that interactions between $\chi$ and the SM always
involve the Yukawa couplings $y_\chi$, $y_\chi'$.  There is no {\it a priori}
constraint on the magnitude of these couplings.  We focus here on
the region of small $y_\chi,\, y_\chi' \lesssim 10^{-7}$, where $\chi$ never
comes into thermal equilibrium with the SM through the interactions in
\cref{eq:Lag}.  We thus imagine that a
thermal abundance of DM is produced during reheating after inflation or by
other new interactions at scales far above the electroweak scale. $\chi$
freezes out when these interactions decouple, at a temperature $T_\text{fo} \gg
m_\chi$. Its abundance at freeze-out is then several orders of magnitude larger
than the value observed today, and is independent of $m_\chi$, $y_\chi$, and
$y_\chi'$.  After freeze-out, the DM number density tracks the entropy density
until $S_3$ develops a vev thanks to the flip-flopping vevs and $\chi$ begins
to decay.  Recall that during that epoch,
$S_3$, $\Psi_3$, and $\Psi_3'$ are still in the thermal bath, so processes with
these particles in the final state are as good at depleting $\chi$ as
processes with only SM particles in the final state would be.

Note that inverse decays like $W \Psi_3 \to \chi$ are negligible in
the parameter space of interest to us as $\chi$ is sufficiently heavy that its
abundance is at all times
larger than the equilibrium abundance. This would be different at larger
values of $y_\chi$, $y_\chi'$, where the interplay of decays and inverse decays
would keep $\chi$ in thermal equilibrium until either a conventional
thermal freeze-out happens or the electroweak phase transition switches
off mixing between $\chi$ and $\Psi_3$, $\Psi_3'$. However, in the parameter region
featuring a vev flip-flop, it turns out that neither of these mechanisms can
yield the correct DM relic density.

Some time after the $\mathbb{Z}_3$ symmetry is restored, the annihilation
processes that keep $S_3$, $\Psi_3$, and $\Psi_3'$ in thermal equilibrium
freeze-out, leaving behind a small, subdominant, relic abundance
of $S_3$, $\Psi_3$, and $\Psi_3'$.  It is crucial in this
context that the electrically neutral components of these fields must be lighter
than the charged components today to avoid stable charged relics in the Universe.
Radiative electroweak corrections indeed lead to a mass splitting
of $\mathcal{O}(100\,\text{MeV})$ \cite{Cirelli:2009uv}.  For $S_3$,
we have to make the additional assumption that $\beta \lesssim 0.001$
to avoid a tree level mass splitting that would shift $s^+$
($s^-$) downward (upward) by
$\tfrac{1}{8} \beta v^2 / m_s$~\cite{Kopp:2013mi}.


\begin{figure}
  \begin{center}
    \includegraphics[width=\columnwidth]{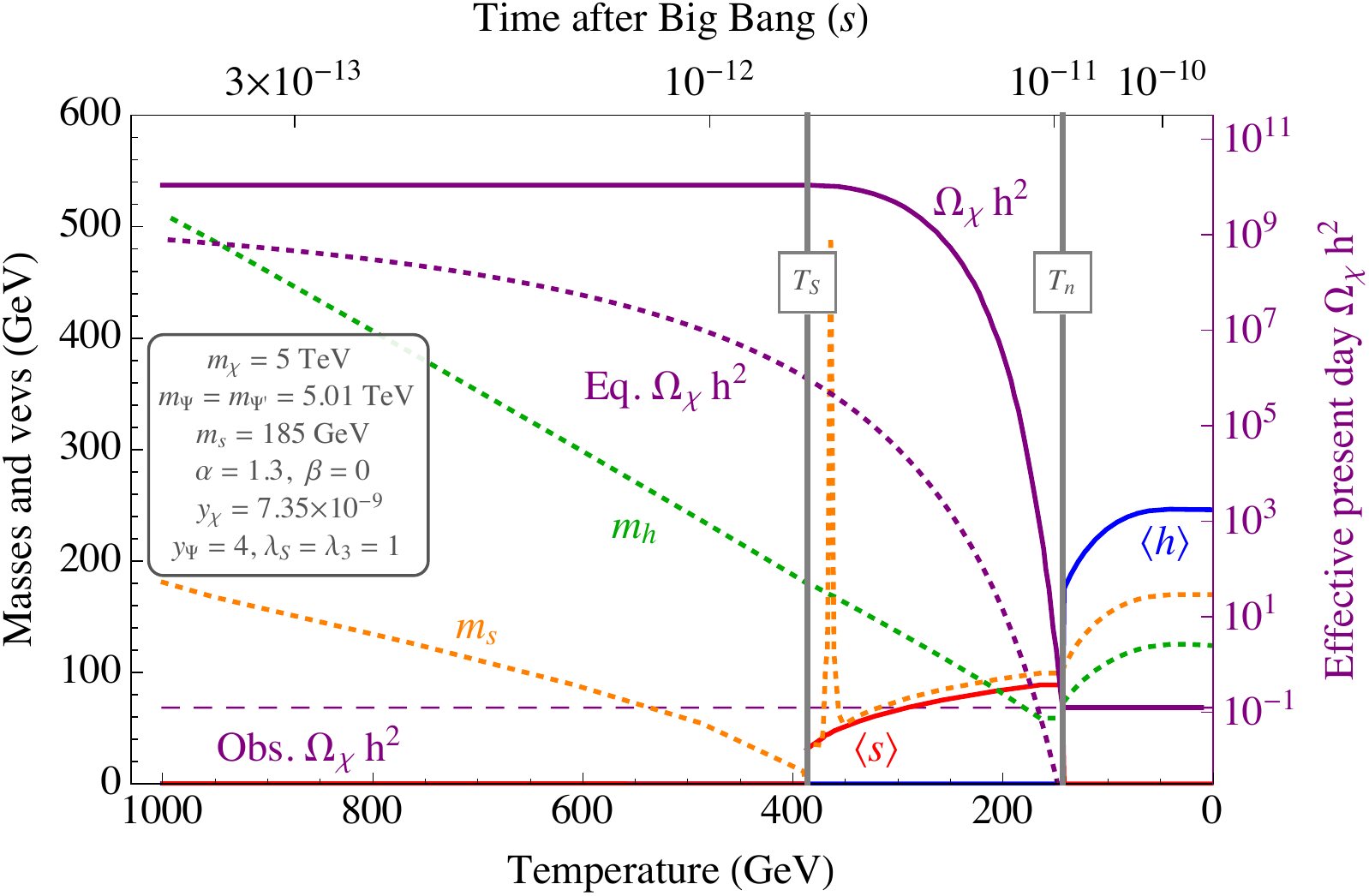}
  \end{center}
  \caption{The vev flip-flop: at very high temperatures, both the SM Higgs
    field vev $\vev{h}$ (solid blue) and the dark sector scalar vev $\vev{s}$
    (solid red) are zero.  As the temperature drops there is a phase transition
    at $T_S$ and $S_3$ develops a nonzero vev, breaking the dark sector
    $\mathbb{Z}_3$ symmetry and making the DM unstable.  We plot in solid
    purple the evolution of the effective present day relic density, obtained
    from the instantaneous DM number density scaled by the subsequent expansion
    of the Universe.  At a lower temperature $T_n$, there is a second phase transition
    where $h$ develops a vev and the feedback of this vev into the dark
    sector via the Higgs portal restores the $\mathbb{Z}_3$ symmetry, halting
    the DM decay when the observed relic density is reached.
    The DM equilibrium density (dotted purple), observed relic density (dashed 
    purple), SM Higgs mass $m_h$ (dotted green),
    and $s$ mass $m_s$ (dotted orange) are also given.
  }
  \label{fig:vev-flip-flop}
\end{figure}


In \cref{fig:vev-flip-flop} we show the evolution of the scalar vevs, the
scalar masses and the effective present day DM relic density as the Universe
cools, for an
illustrative benchmark point with $m_\chi = 5$\,TeV $m_\Psi = m_{\Psi'} =
5.01$\,TeV, and $y_\chi = y_\chi' = 7.35\times10^{-9}$.\footnote{Note that such extreme tuning between $m_\chi$ and $m_\Psi$,
$m_\Psi'$ is not necessary. The desired phenomenology is achieved as
long as $|m_\Psi^{(\prime)} - m_\chi| \lesssim m_S$.}
The parameters of the scalar potential are the same as in
\cref{fig:eff-pot}.  At the left edge of the plot, at $T = 1$\,TeV, both scalar
fields are without vevs, and the $\chi$ relic density is around eleven orders of
magnitude too large.  When the temperature reaches $T_S \sim 385$\,GeV the
Universe transitions (via a first order phase transition) to a phase where
$\vev{s} \neq 0$.  At this point $\chi$ begins to decay.
Most $\chi$ decays happen at temperatures just
above the electroweak phase transition as the temperature drops more slowly at
later times.
At $T \sim 152$\,GeV the $\vev{s} = 0$, $\vev{h} \neq 0$ phase becomes the
global minimum, but as there is a barrier in the potential the Universe does
not immediately transition to this phase; there is a short period of
supercooling until the Universe nucleates to the $\vev{s} = 0$,
$\vev{h} \neq 0$ phase at $T_n \sim 143$\,GeV with another first order phase
transition.  At this point the $\mathbb{Z}_3$ symmetry is restored and DM decays
are no longer possible.

\Cref{fig:vev-flip-flop} also illustrates that the DM abundance today depends
sensitively on the precise value of $T_n$ and thus on the parameters of the
model.  This scenario therefore does not solve the coincidence problem between the
observed baryon and DM abundances. On the other hand, if this model is realised in
nature,
its parameters could be inferred with high precision from cosmological
measurements.

In computing the $\chi$ abundance today, we take into account the momentum
dependence of the decay rate.  At freeze-out, $\chi$ had a Fermi-Dirac distribution,
but because of relativistic time dilation, the low-momentum modes decay faster
than the high-momentum ones, which leads to a skewed distribution at
lower temperatures.  Details on our calculation of the DM relic abundance are
given in appendix~C.
\newline


\begin{figure}
  \begin{center}
    \includegraphics[width=\columnwidth]{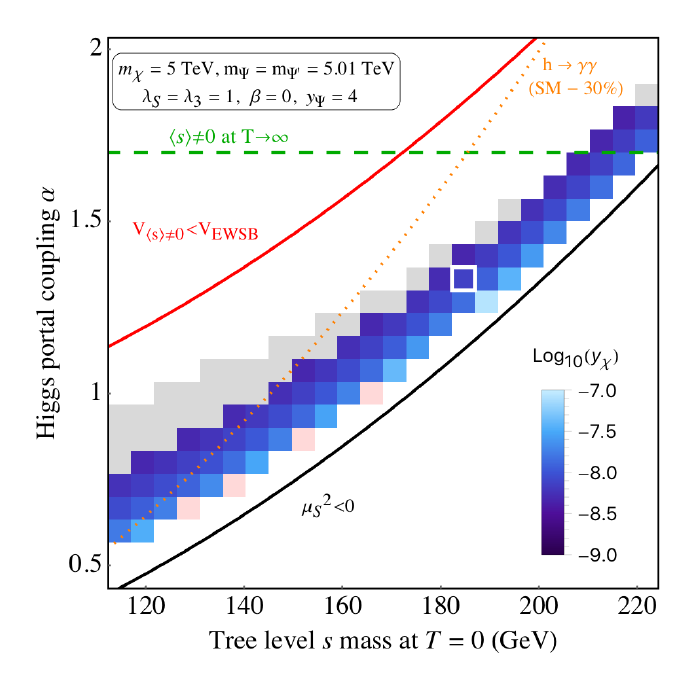}
  \end{center}
  \caption{The value of $y_\chi = y_\chi'$ required to obtain the correct relic density
    $\Omega h^2$~\cite{Ade:2015xua} for DM mass $m_\chi = 5$\,TeV and
    $m_\Psi = m_{\Psi'} = 5.01$\,TeV.  For points 
    plotted in grey, the vev flip-flop occurs but DM depletion is halted
    by freeze-out of $\Psi_3$ and $\Psi_3'$ rather than the electroweak phase transition.
    At points shown in pink the equilibrium abundance
    of $\chi$ is still too high at $T_n$, and as inverse decays ensure that
    the DM density can never drop below its equilibrium value, this implies
    DM overproduction.
    Outside the pixelated region, the vev flip-flop does not occur.
    The red and black lines give tree
    level estimates for the boundaries of this region.
    Above the dotted orange line, the decay rate for $h \to \gamma\gamma$
    is lower than the SM prediction by more than 30\%.
    Above the dashed green line the $\mathbb{Z}_3$
    symmetry is broken in the high $T$ limit. 
  }
  \label{fig:param-space}
\end{figure}


\emph{Parameter Space.}---We now depart from the particular parameter point
considered in \cref{fig:eff-pot,fig:vev-flip-flop} and consider a wider range of
parameter space in \cref{fig:param-space}, which shows the $m_s$--$\alpha$
plane.  Here, $m_s$ refers to
the tree level $s$ mass at $T=0$.  Note that for $\alpha \gtrsim 5$ the
perturbative expansion may no longer be reliable~\cite{Curtin:2014jma,
Tamarit:2014dua}, but for the $\alpha$ values shown
in~\cref{fig:param-space} there is no problem.

We highlight the region where the observed relic density can be obtained from
the vev flip-flop, with the blue colour-code indicating the requisite value of
$y_\chi = y_\chi'$.  We see that this is usually possible provided the
vev flip-flop occurs at all (pixelated area in \cref{fig:param-space}).
At parameter points shown in grey, the $\Psi_3$, $\Psi_3'$ annihilation rate
at the nucleation temperature
$\sim (y_\Psi^4 / m_\Psi^2) [m_\Psi T_n/(2\pi)]^{3/2} \exp(-m_\Psi/T_n)$
is lower than the Hubble rate $\sim T_n^2 / M_\text{Pl}$, implying that
the depletion of dark sector particles is halted by $\Psi_3$, $\Psi_3'$
rather than the electroweak phase transition.
This can change if the DM mass is lower, $y_\Psi$ is larger, or extra annihilation modes
for $\Psi_3$ and $\Psi_3'$ are introduced. Points with too high $T_n$ (shown in pink),
could be rescued by increasing $m_\chi$.

Below the black line in \cref{fig:param-space}, $\mu_S^2$ is negative at tree
level, so the $\mathbb{Z}_3$ symmetry is never broken at tree-level.  Above the
red line, $\mu_S^2 > 0$ is so large at tree level that the Universe never
leaves the $\vev{s} \neq 0$, $\vev{h} = 0$ phase.  By comparing with the edge of the
pixelated region, where the vev flip-flop occurs, we see that these tree level 
estimates are only rough approximations
to the full one-loop $T$ dependent behaviour.  Above the dashed green line,
our model is in the phase with broken $\mathbb{Z}_3$ and
$SU(2)_L$ symmetries in the $T \to \infty$ limit.
In this regime, the positive daisy corrections to
$\mu_S^2$ are larger than the negative one-loop corrections in the high-$T$
limit. In principle, $\chi$ decays are then sensitive to higher scale
physics, but since most decays still happen just above the electroweak phase
transition, our calculations, which begin at $T = 1$\,TeV, remain valid.
\newline


\emph{Constraints and Future Tests.}---The flip-flopping vevs model presented
here is mainly constrained by collider searches as direct and indirect
detection of $\chi$ are hindered by the smallness of $y_\chi$, $y_\chi'$.  For
direct and indirect searches, the subdominant population of $S_3$, $\Psi_3$,
$\Psi_3'$ offers the best detection prospects.

At the LHC or a possible future collider, the most promising direct discovery channel is
Drell--Yan production of $s^\pm$ pairs through their electromagnetic
interactions, followed by the decay $s^\pm \to s + W^{*\pm}$.  Due to the
small mass splitting between the charged and neutral components of the
triplets, these decays will typically be too soft to be observed directly, so
that the sensitivity to the model will come mainly from mono-$X$ searches.
At present, these searches still fall several
orders of magnitude short of probing the interesting region of parameter
space~\cite{Kopp:2013mi}.  For very small mass splittings $< m_\pi \sim
140$\,MeV, charged track searches could be
sensitive~\cite{Aad:2015uaa,CMS:2016ybj,CMS:2016isf}, but reducing the mass
splitting this far would require tuning between a small non-zero value of the
coupling constant $\beta$ and electroweak radiative corrections.
Perhaps the most promising probe of our vev flip-flop model at
the LHC is the precision measurement of the $h \to \gamma\gamma$ rate.
Indeed, as shown in \cref{fig:param-space}, sizeable deviations from
the SM rate for this decay are predicted. In fact, the present limit
would already appear to be in some tension with our model at the $2\sigma$ level
\cite{Khachatryan:2016vau,Olive:2016xmw} due to an event excess in the
7+8\,TeV data from ATLAS and CMS. This statement is, however, put into perspective
when taking into account that at $\sqrt{s} = 13$\,TeV, both
experiments have observed an event deficit~\cite{CMS-PAS-HIG-16-020,
ATLAS-CONF-2016-067}.
\newline


\emph{Summary.}---We have presented a novel mechanism for generating the
observed relic abundance of electroweak scale DM utilising a temporary period
of dark matter decay at the weak scale.  In this scenario, the DM abundance
today is governed by the dynamics of the scalar potential.  We have
demonstrated the mechanism in a simple Higgs portal model, focusing on a
region of parameter space where dark matter freezes out while relativistic.
The initial overabundance is depleted after the $\mathbb{Z}_3$ symmetry that
stabilises dark matter is spontaneously broken and DM decays.
At a later point the Universe undergoes a first order
phase transition that restores the dark sector $\mathbb{Z}_3$ symmetry and
breaks $SU(2)_L \times U(1)_Y$.  Finally, we have outlined avenues for testing this
scenario at colliders.
\newline


\emph{Acknowledgements.}---We would like to thank the members of the DFG
Research Unit ``New Physics at the LHC'', especially Jan Heisig and Susanne
Westhoff, for many useful discussions.  We are moreover indebted to Jonathan
Kozaczuk for invaluable advice on using {\tt CosmoTransitions}.  JK would like
to thank Moshe Moshe for interesting conversations.  We also appreciate suggestions 
made by Tim Cohen, Brian Shuve and Andi Hektor.  We would like to
thank CERN and Fermilab for kind hospitality and support during crucial stages
of this project.  This work was in part supported by the German Research
Foundation (DFG) under Grant Nos.\ \mbox{KO~4820/1--1} and FOR~2239, and by the
European Research Council (ERC) under the European Union's Horizon 2020
research and innovation programme (grant agreement No.\ 637506,
``$\nu$Directions'').
\newline


\appendix
\section{Appendix A: The Scalar Potential at Finite $T$.}

Here we give the finite temperature corrections to the scalar potential in
\cref{eq:Pot}. These corrections are dominated by contributions from the
scalars ($H$ and $S_3$), the electroweak gauge bosons and the top quark. 
The other SM fermions couple very weakly to $H$, while the 
new fermions $\chi$, $\Psi_3$ and $\Psi_3'$ are heavy and decouple at the 
temperatures under consideration. The
field dependent masses of the CP even neutral scalars in the basis
$(h, s)$ are given by
\begin{align}
  \begin{pmatrix}
    -\mu_H^2 + 3 \lambda h^2 + \frac{1}{2}\alpha s^2  &  \alpha h s \\
    \alpha h s   &   -\mu_S^2 + 3 \lambda_S s^2 +\frac{1}{2}\alpha h^2 
  \end{pmatrix}.
  \label{eq:m2-neutral}
\end{align}
\begin{widetext}
The charged scalars, in the basis $(s^+, s^-, G^+)$, have the 
mass matrix
\begin{align}
  \begin{pmatrix}
  -\mu_S^2 + \frac{1}{4}(2\alpha-\beta) h^2 + (\lambda_S + \lambda_3) s^2 &
  \lambda_3 s^2 &
  \frac{\beta h s}{2\sqrt{2}}
  \\
  \lambda_3 s^2 &
  -\mu_S^2 + \frac{1}{4}(2\alpha+\beta) h^2 + (\lambda_S + \lambda_3) s^2 & 
  \frac{\beta h s}{2\sqrt{2}}
  \\
  \frac{\beta h s}{2\sqrt{2}} &
  \frac{\beta h s}{2\sqrt{2}}&
  -\mu_H^2 + \lambda_H h^2 + \frac{\alpha s^2}{2}
  \end{pmatrix}.
\end{align}
\end{widetext}
The field dependent masses of the remaining fields are
\begin{align}
  m_a^2       &= -\mu_S^2 + \frac{1}{2} \alpha h^2 + \lambda_S s^2 \,,\\
  m_{G^0}^2   &= -\mu_H^2 + \frac{1}{2} \alpha s^2 + \lambda_H h^2 \,,\\
  m_{W^\pm}^2 &= \frac{1}{4} g^2 (h^2 + 4 s^2) \,,\\
  m_{Z}^2     &= \frac{1}{4}(g^2 + g'^2) h^2 \,,\\
  m_\gamma^2  &= 0 \,, \label{eq:m2-gamma} \\
  m_t^2       &= \frac{1}{2} y_t^2 h^2 \,.
\end{align}
Here, $y_t$ is the top quark Yukawa coupling and $g$ and $g'$ are the SM
$SU(2)_L$ and $U(1)_Y$ coupling constants, respectively.

The $T$-independent Coleman-Weinberg contribution
is~\cite{Coleman:1973jx, Quiros:1999jp}
\begin{align}
  \hspace*{-0.5cm}
  V^\text{CW}(h,s) &=
    \sum_i \frac{n_i}{64 \pi^2} m_i^4(h,s)
    \left[ \log
      \frac{m_i^2(h,s)}{\Lambda^2} - C_i \right],
\end{align}
where the sum is over the eigenvalues of the mass matrices of 
$\lbrace h, G^0, G^+, s, a, s^+, s^-, W, Z, t
\rbrace$ and $n_h = n_s = n_a = n_{G^0} = 1$, $n_{G^+} = n_{s^\pm} =
2$, $n_Z = 3$, $n_W = 6$ and $n_t=-12$ accounts for their degrees of freedom. 
We take the renormalisation scale $\Lambda$ to be the 
SM Higgs vev $v = 246$\,GeV.  In the dimensional regularisation scheme 
$C_i = 5/6 \,(3/2)$ for gauge bosons (scalars and fermions).  We also 
add counterterms to ensure that $v = \mu_H/\sqrt{\lambda_H}$ and $m_h = \sqrt{2}\mu_H$
at $T=0$.

The one-loop finite temperature correction is~\cite{Dolan:1973qd}
\begin{align}
  V^T(h,s) =& \sum_i \frac{n_i T^4}{2\pi^2}\int_0^\infty \! dx \, x^2
                                                                \nonumber\\[0.2cm]
    &\hspace{-1.5cm} \times \log \bigg[ 1 \pm
\exp\Big(-\sqrt{x^2+m_i^2(h,s)/T^2}\Big)\bigg] \,,
\end{align}
where we sum over the same eigenvalues as for the Coleman-Weinberg contribution. 
The negative sign in the integrand is for bosons while the positive sign is
for fermions.

The bosons also contribute to higher order ``daisy'' diagrams which can be
resummed to give~\cite{Carrington:1991hz}
\begin{align}
  V^\text{daisy} &= -\frac{T}{12\pi} \sum_i n_i
    \Big( \left[ m^2(h,s) + \Pi(T) \right]_i^\frac{3}{2}  \nonumber\\[0.2cm] 
  &\hspace{3.0cm}
    - \left[ m_i^2(h,s) \right]_i^\frac{3}{2}
  \Big) \,.
\end{align}
Here, the first term should be interpreted as the $i$-th eigenvalue of the
matrix-valued quantity $[m^2(h,s) + \Pi(T)]^{3/2}$, where $m^2(h,s)$ is the
block-diagonal matrix composed of the individual mass matrices in
\crefrange{eq:m2-neutral}{eq:m2-gamma}~\cite{Patel:2011th}.
The sum runs over the bosonic degrees of freedom:
$i \in \lbrace h, G^0, G^+, s, a, s^+, s^-, W^i, B \rbrace$.
The bosonic thermal masses (Debye masses) in the gauge basis
are given by
\begin{align}
  \Pi_{h,G^0, G^+}    &= \frac{T^2}{16}\left(3 g^2 + g'^2 + 4\alpha
                       + 8\lambda_H + 4 y_t^2\right) \,,\\
  \Pi_{s,a, s^+, s^-} &= \frac{T^2}{6}\left(3 g^2 +  \alpha
                       + 2\lambda_3 + 4\lambda_S \right) \,,\\
  \Pi_{W^{1,2,3}}^L   &= \frac{5}{2}g^2T^2 \,
     \qquad
  \Pi_{W^{1,2,3}}^T = 0 \,,\\
  \Pi_B^L             &= \frac{11}{6}g'^2T^2 \,,
     \qquad
  \Pi_B^T = 0.
\end{align}
Since only the
longitudinal components of the gauge bosons contribute, $n_{W^i}^L = n_B^L= 1$.

\section{Appendix B: Dark Matter Decay Rates.}

When $\ev{s} \neq 0$, the fermion $\chi$ can mix into $\Psi_3$ and $\Psi_3'$
and decay via a $W$ boson.  When kinematically allowed, $\chi$ can also decay
into components of $\Psi_3$ or $\Psi_3'$ and a component of $S_3$.  We will
denote the mass eigenstates of the charged fermions as
$\widetilde{\Psi}^+_{1,2}$, and we will make use of the functions
\begin{widetext}
\begin{align}
  \Gamma_V(W, \psi)
    &= \frac{\sqrt{[m_\chi^2 - (m_W - m_\psi)^2]
                   [m_\chi^2 - (m_W + m_\psi)^2]}}
            {16 \pi m_W^2 m_{\chi}^3}
       \Big[-2 m_W^4 + m_W^2 (m_\chi^2 - 6 m_\chi m_\psi+ m_\psi^2)
          + (m_\chi^2 - m_\psi^2)^2 \Big] \,,\\
  \Gamma_S(s,\psi)
    &= \frac{\sqrt{[m_\chi^2 - (m_s - m_{\psi})^2]
                   [m_\chi^2 - (m_s + m_{\psi})^2]}}
            {16 \pi  m_\chi^3}
       \Big[ (m_\chi + m_\psi)^2 - m_s^2 \Big] \,.
\end{align}
\end{widetext}

Taking $y_\chi = y_\chi'$, we find the total decay rate to be
\begin{align}
  \Gamma_\chi
    &= g^2 \sin^2 \theta \big[ \Gamma_V(W, \widetilde{\psi}^+_1)
                             + \Gamma_V(W, \widetilde{\psi}^+_2) \big]
             + y_\chi^2 \Gamma_S(s, \psi^0) \notag\\
    &+ \left( y_\chi^2 + \frac{y_\Psi^2 \sin^2\theta}{2}\right)
       \big[ \Gamma_S(s^+, \widetilde{\Psi}^+_1)
           + \Gamma_S(s^+, \widetilde{\Psi}^+_2) \big] \,.
\end{align}
The mixing angle is $\theta \simeq y_\chi \ev{s}/(m_\Psi - m_\chi)$, where we
have used the small angle approximation.  This is justified
since $y_\chi \lesssim 10^{-7}$ and $\ev{s} \sim (m_\Psi - m_\chi)$. Since the mixing
angle is small, we neglect the resulting change in the $\chi$ and $\Psi^0$
mass eigenvalues.

\section{Appendix C: Computation of the Dark Matter Abundance.}

To compute the DM abundance today in the vev flip-flop scenario, we need to
solve the coupled Boltzmann equations
\begin{align}
  \dot{n}_\chi^j + 3 H n_\chi^j
    = -\frac{\Gamma}{\gamma^j} \big( n_\chi^j - n_\chi^{j,\text{eq}} \big)
  \label{eq:Boltzmann}
\end{align}
together with the Friedmann equation
\begin{align}
  H^2 = \bigg( \frac{\dot{a}}{a} \bigg)^2
    = \frac{8\pi G_N}{3} \big( \rho_\text{SM} + \rho_\chi \big) \,.
  \label{eq:Friedmann}
\end{align}
In these equations, $n_\chi^j$ denotes the number density of DM particles in a
comoving momentum interval centred at a momentum $p^j$, and
$n_\chi^{j,\text{eq}}$ is the corresponding equilibrium abundance. We use 128
such intervals, covering the momentum range from 0 to 1000\,TeV at $T=10$\,TeV.
We have checked that increasing the momentum range or resolution does not alter
our results.  The DM energy density is obtained from $n_\chi^j$ according to
$\rho_\chi \equiv \sum_j E^j n_\chi^j$, where $E^j = [(p^j)^2 +
m_\chi^2]^{1/2}$, and the energy density of SM particles is given by
$\rho_\text{SM} = g_* (\pi^2 / 30) T^4$, with $g_*$ the effective number of SM
degrees of freedom.  The decay rate $\Gamma$ depends on temperature through
$m_s$, $m_h$ and $\vev{s}$, but is independent of momentum. The relativistic
gamma factor is given by $\gamma^j = E_j / m_\chi$.

To solve \cref{eq:Boltzmann,eq:Friedmann} in practice, we make the substitution
$Y^j \equiv n^j / s$, where $s = g_* (2 \pi^2 / 45) T^3$ is the entropy density
in SM degrees of freedom.  We moreover need a relation between time $t$ and
temperature $T$. To find it, note that
\begin{align}
  \dot\rho_\text{SM}
  = 4 \rho_\text{SM} \frac{\dot{T}}{T}
  = m_\chi \sum_j \frac{\Gamma}{\gamma^j} \big( n_\chi^j - n_\chi^{j,\text{eq}} \big)
  - 4 H \rho_\text{SM} \,.
  \label{eq:rho-SM-dot}
\end{align}
The first term on the right hand side corresponds to energy injection into
the SM plasma from DM decays, the second one corresponds to energy dissipation
by redshifting.  It follows that
\begin{align}
  \frac{dT}{dt}
    = \sum_j \big( n_\chi^j - n_\chi^{j,\text{eq}} \big)
      \frac{m_\chi \Gamma T}{4 \gamma^j \rho_\text{SM}}
    - H T \,.
  \label{eq:dT-dt}
\end{align}

\bibliographystyle{JHEP}
\bibliography{./interacting-vevs}

\providecommand{\href}[2]{#2}\begingroup\raggedright\begin{thebibliography}{10}

\bibitem{LUX:2016}
{\bf LUX} Collaboration, A.~Manalaysay, {\it {Dark-matter results from 332 new
  live days of LUX data}}, . Identification of Dark Matter 2016
  \href{http://lux.brown.edu/LUX_dark_matter/Talks_files/LUX_NewDarkMatterSearchResult_332LiveDays_IDM2016_160721.pdf}{slides}.

\bibitem{Tan:2016zwf}
{\bf PandaX-II} Collaboration, A.~Tan et~al., {\it {Dark Matter Results from
  First 98.7-day Data of PandaX-II Experiment}},
  \href{http://arxiv.org/abs/1607.07400}{{\tt arXiv:1607.07400}}.

\bibitem{Ackermann:2015zua}
{\bf Fermi-LAT} Collaboration, M.~Ackermann et~al., {\it {Searching for Dark
  Matter Annihilation from Milky Way Dwarf Spheroidal Galaxies with Six Years
  of Fermi Large Area Telescope Data}},  {\em Phys. Rev. Lett.} {\bf 115}
  (2015), no.~23 231301, [\href{http://arxiv.org/abs/1503.02641}{{\tt
  arXiv:1503.02641}}].

\bibitem{Pizzolotto:2016lwk}
{\bf AMS-02} Collaboration, C.~Pizzolotto, {\it {Positron fraction, electron
  and positron spectra measured by AMS-02}},  {\em EPJ Web Conf.} {\bf 121}
  (2016) 03006.

\bibitem{Madhavacheril:2013cna}
M.~S. Madhavacheril, N.~Sehgal, and T.~R. Slatyer, {\it {Current Dark Matter
  Annihilation Constraints from CMB and Low-Redshift Data}},  {\em Phys. Rev.}
  {\bf D89} (2014) 103508, [\href{http://arxiv.org/abs/1310.3815}{{\tt
  arXiv:1310.3815}}].

\bibitem{Aaboud:2016tnv}
{\bf ATLAS} Collaboration, M.~Aaboud et~al., {\it {Search for new phenomena in
  final states with an energetic jet and large missing transverse momentum in
  $pp$ collisions at $\sqrt{s}=13$ TeV using the ATLAS detector}},
  \href{http://arxiv.org/abs/1604.07773}{{\tt arXiv:1604.07773}}.

\bibitem{CMS:2016pod}
{\bf CMS} Collaboration, {\it {Search for dark matter in final states with an
  energetic jet, or a hadronically decaying W or Z boson using
  $12.9~\mathrm{fb}^{-1}$ of data at $\sqrt{s} = 13~\mathrm{TeV}$}}, .
  \href{https://cds.cern.ch/record/2205746}{CMS-PAS-EXO-16-037}.

\bibitem{Profumo:2007wc}
S.~Profumo, M.~J. Ramsey-Musolf, and G.~Shaughnessy, {\it {Singlet Higgs
  phenomenology and the electroweak phase transition}},  {\em JHEP} {\bf 08}
  (2007) 010, [\href{http://arxiv.org/abs/0705.2425}{{\tt arXiv:0705.2425}}].

\bibitem{Cline:2009sn}
J.~M. Cline, G.~Laporte, H.~Yamashita, and S.~Kraml, {\it {Electroweak Phase
  Transition and LHC Signatures in the Singlet Majoron Model}},  {\em JHEP}
  {\bf 07} (2009) 040, [\href{http://arxiv.org/abs/0905.2559}{{\tt
  arXiv:0905.2559}}].

\bibitem{Espinosa:2011ax}
J.~R. Espinosa, T.~Konstandin, and F.~Riva, {\it {Strong Electroweak Phase
  Transitions in the Standard Model with a Singlet}},  {\em Nucl. Phys.} {\bf
  B854} (2012) 592--630, [\href{http://arxiv.org/abs/1107.5441}{{\tt
  arXiv:1107.5441}}].

\bibitem{Cui:2011qe}
Y.~Cui, L.~Randall, and B.~Shuve, {\it {Emergent Dark Matter, Baryon, and
  Lepton Numbers}},  {\em JHEP} {\bf 08} (2011) 073,
  [\href{http://arxiv.org/abs/1106.4834}{{\tt arXiv:1106.4834}}].

\bibitem{Cline:2012hg}
J.~M. Cline and K.~Kainulainen, {\it {Electroweak baryogenesis and dark matter
  from a singlet Higgs}},  {\em JCAP} {\bf 1301} (2013) 012,
  [\href{http://arxiv.org/abs/1210.4196}{{\tt arXiv:1210.4196}}].

\bibitem{Fairbairn:2013uta}
M.~Fairbairn and R.~Hogan, {\it {Singlet Fermionic Dark Matter and the
  Electroweak Phase Transition}},  {\em JHEP} {\bf 09} (2013) 022,
  [\href{http://arxiv.org/abs/1305.3452}{{\tt arXiv:1305.3452}}].

\bibitem{Curtin:2014jma}
D.~Curtin, P.~Meade, and C.-T. Yu, {\it {Testing Electroweak Baryogenesis with
  Future Colliders}},  {\em JHEP} {\bf 11} (2014) 127,
  [\href{http://arxiv.org/abs/1409.0005}{{\tt arXiv:1409.0005}}].

\bibitem{Cohen:2008nb}
T.~Cohen, D.~E. Morrissey, and A.~Pierce, {\it {Changes in Dark Matter
  Properties After Freeze-Out}},  {\em Phys. Rev.} {\bf D78} (2008) 111701,
  [\href{http://arxiv.org/abs/0808.3994}{{\tt arXiv:0808.3994}}].

\bibitem{Coleman:1973jx}
S.~R. Coleman and E.~J. Weinberg, {\it {Radiative Corrections as the Origin of
  Spontaneous Symmetry Breaking}},  {\em Phys. Rev.} {\bf D7} (1973)
  1888--1910.

\bibitem{Dolan:1973qd}
L.~Dolan and R.~Jackiw, {\it {Symmetry Behavior at Finite Temperature}},  {\em
  Phys. Rev.} {\bf D9} (1974) 3320--3341.

\bibitem{Carrington:1991hz}
M.~E. Carrington, {\it {The Effective potential at finite temperature in the
  Standard Model}},  {\em Phys. Rev.} {\bf D45} (1992) 2933--2944.

\bibitem{Quiros:1999jp}
M.~Quiros, {\it {Finite temperature field theory and phase transitions}},  in
  {\em {Proceedings, Summer School in High-energy physics and cosmology:
  Trieste, Italy, June 29-July 17, 1998}}, pp.~187--259, 1999.
\newblock \href{http://arxiv.org/abs/hep-ph/9901312}{{\tt hep-ph/9901312}}.

\bibitem{Ahriche:2007jp}
A.~Ahriche, {\it {What is the criterion for a strong first order electroweak
  phase transition in singlet models?}},  {\em Phys. Rev.} {\bf D75} (2007)
  083522, [\href{http://arxiv.org/abs/hep-ph/0701192}{{\tt hep-ph/0701192}}].

\bibitem{Delaunay:2007wb}
C.~Delaunay, C.~Grojean, and J.~D. Wells, {\it {Dynamics of Non-renormalizable
  Electroweak Symmetry Breaking}},  {\em JHEP} {\bf 04} (2008) 029,
  [\href{http://arxiv.org/abs/0711.2511}{{\tt arXiv:0711.2511}}].

\bibitem{Linde:1981zj}
A.~D. Linde, {\it {Decay of the False Vacuum at Finite Temperature}},  {\em
  Nucl. Phys.} {\bf B216} (1983) 421. [Erratum: Nucl. Phys.B223,544(1983)].

\bibitem{Wainwright:2011kj}
C.~L. Wainwright, {\it {CosmoTransitions: Computing Cosmological Phase
  Transition Temperatures and Bubble Profiles with Multiple Fields}},  {\em
  Comput. Phys. Commun.} {\bf 183} (2012) 2006--2013,
  [\href{http://arxiv.org/abs/1109.4189}{{\tt arXiv:1109.4189}}].

\bibitem{Kozaczuk:2014kva}
J.~Kozaczuk, S.~Profumo, L.~S. Haskins, and C.~L. Wainwright, {\it
  {Cosmological Phase Transitions and their Properties in the NMSSM}},  {\em
  JHEP} {\bf 01} (2015) 144, [\href{http://arxiv.org/abs/1407.4134}{{\tt
  arXiv:1407.4134}}].

\bibitem{Blinov:2015sna}
N.~Blinov, J.~Kozaczuk, D.~E. Morrissey, and C.~Tamarit, {\it {Electroweak
  Baryogenesis from Exotic Electroweak Symmetry Breaking}},  {\em Phys. Rev.}
  {\bf D92} (2015), no.~3 035012, [\href{http://arxiv.org/abs/1504.05195}{{\tt
  arXiv:1504.05195}}].

\bibitem{Kozaczuk:2015owa}
J.~Kozaczuk, {\it {Bubble Expansion and the Viability of Singlet-Driven
  Electroweak Baryogenesis}},  {\em JHEP} {\bf 10} (2015) 135,
  [\href{http://arxiv.org/abs/1506.04741}{{\tt arXiv:1506.04741}}].

\bibitem{Patel:2011th}
H.~H. Patel and M.~J. Ramsey-Musolf, {\it {Baryon Washout, Electroweak Phase
  Transition, and Perturbation Theory}},  {\em JHEP} {\bf 07} (2011) 029,
  [\href{http://arxiv.org/abs/1101.4665}{{\tt arXiv:1101.4665}}].

\bibitem{Cirelli:2009uv}
M.~Cirelli and A.~Strumia, {\it {Minimal Dark Matter: Model and results}},
  {\em New J.Phys.} {\bf 11} (2009) 105005,
  [\href{http://arxiv.org/abs/0903.3381}{{\tt arXiv:0903.3381}}].

\bibitem{Kopp:2013mi}
J.~Kopp, E.~T. Neil, R.~Primulando, and J.~Zupan, {\it {From gamma ray line
  signals of dark matter to the LHC}},  {\em Phys.Dark Univ.} {\bf 2} (2013)
  22--34, [\href{http://arxiv.org/abs/1301.1683}{{\tt arXiv:1301.1683}}].

\bibitem{Ade:2015xua}
{\bf Planck} Collaboration, P.~A.~R. Ade et~al., {\it {Planck 2015 results.
  XIII. Cosmological parameters}},  {\em Astron. Astrophys.} {\bf 594} (2016)
  A13, [\href{http://arxiv.org/abs/1502.01589}{{\tt arXiv:1502.01589}}].

\bibitem{Tamarit:2014dua}
C.~Tamarit, {\it {Higgs vacua with potential barriers}},  {\em Phys. Rev.} {\bf
  D90} (2014), no.~5 055024, [\href{http://arxiv.org/abs/1404.7673}{{\tt
  arXiv:1404.7673}}].

\bibitem{Kuzmin:1985mm}
V.~A. Kuzmin, V.~A. Rubakov, and M.~E. Shaposhnikov, {\it {On the Anomalous
  Electroweak Baryon Number Nonconservation in the Early Universe}},  {\em
  Phys. Lett.} {\bf B155} (1985) 36.

\bibitem{Shaposhnikov:1986jp}
M.~E. Shaposhnikov, {\it {Possible Appearance of the Baryon Asymmetry of the
  Universe in an Electroweak Theory}},  {\em JETP Lett.} {\bf 44} (1986)
  465--468. [Pisma Zh. Eksp. Teor. Fiz.44,364(1986)].

\bibitem{Shaposhnikov:1987tw}
M.~E. Shaposhnikov, {\it {Baryon Asymmetry of the Universe in Standard
  Electroweak Theory}},  {\em Nucl. Phys.} {\bf B287} (1987) 757--775.

\bibitem{Morrissey:2012db}
D.~E. Morrissey and M.~J. Ramsey-Musolf, {\it {Electroweak baryogenesis}},
  {\em New J. Phys.} {\bf 14} (2012) 125003,
  [\href{http://arxiv.org/abs/1206.2942}{{\tt arXiv:1206.2942}}].

\bibitem{Carena:1996wj}
M.~Carena, M.~Quiros, and C.~E.~M. Wagner, {\it {Opening the window for
  electroweak baryogenesis}},  {\em Phys. Lett.} {\bf B380} (1996) 81--91,
  [\href{http://arxiv.org/abs/hep-ph/9603420}{{\tt hep-ph/9603420}}].

\bibitem{Aad:2015uaa}
{\bf ATLAS} Collaboration, G.~Aad et~al., {\it {Search for long-lived, weakly
  interacting particles that decay to displaced hadronic jets in proton-proton
  collisions at $\sqrt{s}=8$ TeV with the ATLAS detector}},  {\em Phys. Rev.}
  {\bf D92} (2015), no.~1 012010, [\href{http://arxiv.org/abs/1504.03634}{{\tt
  arXiv:1504.03634}}].

\bibitem{CMS:2016ybj}
{\bf CMS} Collaboration, {\it {Search for heavy stable charged particles with
  $12.9~\mathrm{fb}^{-1}$ of 2016 data}}, .
  \href{http://cms-results.web.cern.ch/cms-results/public-results/preliminary-results/EXO-16-036/index.html}{CMS-PAS-EXO-16-036}.

\bibitem{CMS:2016isf}
{\bf CMS} Collaboration, {\it {Search for displaced leptons in the e-mu
  channel}}, . \href{https://cds.cern.ch/record/2205146}{CMS-PAS-EXO-16-022}.

\bibitem{Khachatryan:2016vau}
{\bf ATLAS, CMS} Collaboration, G.~Aad et~al., {\it {Measurements of the Higgs
  boson production and decay rates and constraints on its couplings from a
  combined ATLAS and CMS analysis of the LHC pp collision data at $ \sqrt{s}=7
  $ and 8 TeV}},  {\em JHEP} {\bf 08} (2016) 045,
  [\href{http://arxiv.org/abs/1606.02266}{{\tt arXiv:1606.02266}}].

\bibitem{Olive:2016xmw}
{\bf Particle Data Group} Collaboration, C.~Patrignani et~al., {\it {Review of
  Particle Physics}},  {\em Chin. Phys.} {\bf C40} (2016), no.~10 100001.

\bibitem{CMS-PAS-HIG-16-020}
{\bf CMS Collaboration} Collaboration, {\it {Updated measurements of Higgs
  boson production in the diphoton decay channel at $\sqrt{s}=13~\textrm{TeV}$
  in pp collisions at CMS.}},  Tech. Rep. CMS-PAS-HIG-16-020, CERN, Geneva,
  2016.

\bibitem{ATLAS-CONF-2016-067}
{\bf ATLAS Collaboration} Collaboration, {\it {Measurement of fiducial,
  differential and production cross sections in the $H\to\gamma\gamma$ decay
  channel with 13.3 fb$^{-1}$ of 13 TeV proton-proton collision data with the
  ATLAS detector}},  Tech. Rep. ATLAS-CONF-2016-067, CERN, Geneva, Aug, 2016.

\end{thebibliography}\endgroup

\end{document}